\documentclass[namedreferences]{solarphysics}
\usepackage[hyperref,optionalrh,solaromanenum]{spr-sola-addons} 
\usepackage{graphicx}                    
\usepackage{color}                       
\usepackage{breakurl}                         

\newcommand{\ion}[2]{#1\,{\sc #2}}

\chardef\us=`\_

\begin{document}

\begin{article}

\begin{opening}

\title{Estimate of the Upper Limit on Hot Plasma Differential Emission Measure (DEM) in Non-Flaring Active Regions and Nanoflare Frequency Based on the \ion{Mg}{xii} Spectroheliograph Data from CORONAS-F/SPIRIT}

%
\author[addressref={aff1},corref,email={reva.antoine@gmail.com}]{\inits{A. A.}\fnm{Anton}~\lnm{Reva}\orcid{https://orcid.org/0000-0003-4805-1424}}
\author[addressref={aff1}]{\inits{A. S.}\fnm{Artem}~\lnm{Ulyanov}}
\author[addressref={aff1}]{\inits{A. S.}\fnm{Alexey}~\lnm{Kirichenko}}
\author[addressref={aff1}]{\inits{S. A.}\fnm{Sergey}~\lnm{Bogachev}}
\author[addressref={aff1}]{\inits{S. V.}\fnm{Sergey}~\lnm{Kuzin}}

\runningauthor{A. Reva {\it et al.}}
\runningtitle{Estimate of the Upper Limit on Hot Plasma DEM in Non-Flaring Active Regions}

\address[id={aff1}]{Lebedev Physical Institute of Russian Academy of Sciences, Leninskij Prospekt 53, Moscow 119991, Russia}
           
\begin{abstract}
The nanoflare-heating theory predicts steady hot plasma emission in the non-flaring active regions. It is hard to find this emission with conventional non-monochromatic imagers (such as {\it Atmospheric Imaging Assembly} or {\it X-Ray Telescope}), because their images contain a cool temperature background. In this work, we search for hot plasma in non-flaring active regions using the \ion{Mg}{xii} spectroheliograph onboard {\it Complex Orbital Observations Near-Earth of Activity on the Sun (CORONAS)-F/SPectroheliographIc X-ray Imaging Telescope (SPIRIT)}. This instrument acquired monochromatic images of the solar corona in the \ion{Mg}{xii} 8.42 \AA~line, which emits only at temperatures higher than 4 MK. The \ion{Mg}{xii} images contain the signal only from hot plasma without any low-temperature background. We studied the hot plasma in active regions using the SPIRIT data from 18\,--\,28 February 2002. During this period, the \ion{Mg}{xii} spectroheliograph worked with a 105-second cadence almost without data gaps. The hot plasma was observed only in the flaring active regions. We do not observe any hot plasma in non-flaring active regions. The hot plasma column emission measure in the non-flaring active region should not exceed $3 \times 10^{24}$~cm$^{-5}$. The hot Differential Emission Measure (DEM) is less than 0.01\,\% of the DEM of the main temperature component. Absence of \ion{Mg}{xii} emission in the non-flaring active regions can be explained by weak and frequent nanoflares (delay less than 500 seconds) or by very short and intense nanoflares that lead to non-equilibrium ionization.
\end{abstract}


\end{opening}

\section{Introduction}

The solar corona has a temperature around 1\,MK, which is much higher than the temperature of the photosphere. It is unknown why the corona is so hot. 

Today, there are two main theories of the coronal heating: by nanoflares and by MHD waves. Nanoflare theory assumes that many small-scale flares (nanoflares) occur in the corona \citep{par88, Klimchuk2015}. The nanoflares result from small-scale reconnection episodes. Although the energy of each nanoflare is small, their total energy could be sufficient to heat the corona.

In the wave theory, the photosphere generates MHD waves that propagate to the corona, where they dissipate and heat the corona \citep{Biermann1948, Davila1987, Schwarzschild1948}. Waves can produce high-frequency small-scale events that will be indistinguishable from nanoflares \citep{Klimchuk2006}. For this reason, we will use the term ``nanoflare'' to denote a small-scale impulsive energy release event without regard to its nature (waves or reconnection).

Both nanoflares and waves (that propagate from the photosphere to the corona) are hard to detect directly. That is why the experimental tests of these models are focused on finding ``observables'' that these models predict.

One such observable is faint emission of hot plasma (temperature greater than 4~MK) in non-flaring active regions. If the delay between nanoflares on an individual field line is less than the loop cooling time (high-frequency heating), the loop is heated by weak and frequent nanoflares and its temperature slightly deviates from the average value. If the delay between nanoflares is greater than the loop cooling time (low-frequency heating), the loop is heated by stronger but less-frequent nanoflares. In this case, the loop temperature will significantly deviate from the average value. To maintain  the average coronal temperature, the nanoflare should heat the loop to temperatures greater than 4\,MK. Due to the large number of nanoflares, this will create a steady hot-plasma faint emission \citep{Cargill1994, Cargill2014, Klimchuk2015}. Detection of hot plasma in non-flaring active regions would be an indirect evidence of low-frequency nanoflare heating. Absence of hot emission can help to constrain the nanoflare frequency.

However, it is hard to find faint hot-plasma emission with conventional non-monochromatic imagers--like the {\it Atmospheric Imaging Assembly} \citep[AIA:][]{Lemen2012} or the {\it X-Ray Telescope} \citep[XRT:][]{Golub2007}--because their images contain a cool temperature background. Although some researchers \citep{Reale2011, Warren2012, Testa2012} have developed methods to subtract most of the cold background from AIA images, the resulting data still contain the mixed signal, which comes from the hot and cold plasma. 

\citet{Reale2009}, \citet{Schmelz2009b, Schmelz2009a}, and \citet{Testa2011} reported detection of hot plasma using XRT data. However, \citet{Winebarger2012} showed that the XRT data cannot be used to confirm the existence of hot plasma with low emission measure. Due to the broad temperature-response function of the XRT, the instrument can only detect hot plasma with an emission measure that is higher than 10\,\% of the emission measure of the warm component ($\approx$\,2\,--\,3\,MK).

Several researchers have studied hot plasma in non-flaring active regions using monochromatic observations. In hard X-ray, the observations were made with focusing hard X-ray telescopes: {\it Nuclear Spectroscopic Telescope Array} \citep[NuSTAR:][]{Hannah2016, Harrison2013} and {\it Focusing Optics X-ray Solar Imager} \citep[FOXSI:][]{Ishikawa2014, Krucker2014}. In soft X-ray, non-imaging spectrometers were used: {\it Solar Maximum Mission (SMM)/Flat Crystal Spectrometer (FCS)} \citep{DelZanna2014}, {\it Solar PHotometer IN X-rays} \citep[SphinX:][]{Miceli2012, Gburek2011}, and {\it REntgenovsky Spectrometer s Izognutymi Kristalami} \citep[RESIK:][]{Sylwester2010, Sylwester2005}. In the EUV range, the observations were made with imaging spectrometers: {\it Solar Ultraviolet Measurement of Emitted Radiation} \citep[SUMER:][]{Parenti2017, Wilhelm1995} and {\it Extreme Ultraviolet Normal Incidence Spectrograph} \citep[EUNIS-13:][]{Brosius2014}. These works estimated upper limit on the emission measure of the hot plasma in non-flaring active regions.

To test the nanoflare-heating model, we need some way to measure the quantity of the hot plasma ($\approx$\,10\,MK) relative to warm plasma (temperature at DEM maximum, $\approx$\,3\,MK). At the same time, we need enough images of hot plasma to distinguish non-flaring active regions from flaring ones.

For this purpose, we need at least two imaging instruments. The first one should be sensitive to hot plasma and at the same time blind to cool plasma. The second one should image the cool plasma. The instruments should be cross-calibrated and have a rapid enough cadence to observe the dynamics of the hot plasma.

In this paper, we try to set an upper limit on the hot-plasma differential emission measure (DEM) using direct observations of the hot plasma by the \ion{Mg}{xii} spectroheliograph \citep{zhi03a} onboard {\it Complex Orbital Observations Near-Earth of Activity on the Sun (CORONAS)-F/SPectroheliographIc X-ray Imaging Telescope (SPIRIT)} \citep{ora02, Zhitnik2002}. This instrument imaged coronal hot plasma without low-temperature background. We will compare the obtained limit with the result of recent numerical simulations and will try to put constraints on the parameters of the nanoflare-heating model.

\section{Experimental Data}

In our research, we use the data of the \ion{Mg}{xii} spectroheliograph and the data of the {\it Extreme ultraviolet Imaging Telescope} \citep[EIT:][]{del95}.

\begin{figure}[t]
\centering
\includegraphics[width = 0.70\textwidth]{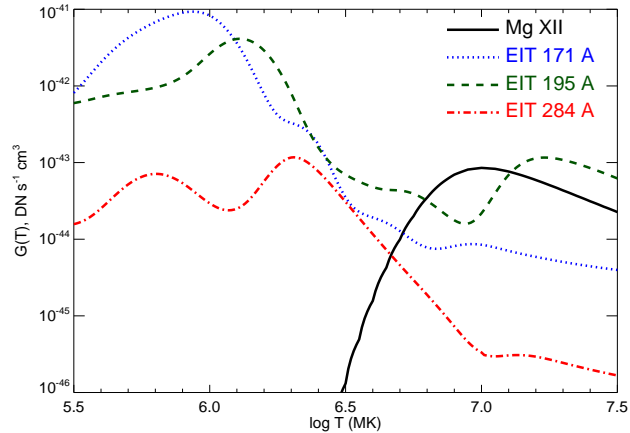}
\caption{Temperature response functions of the \ion{Mg}{xii} spectroheliograph and EIT channels. Black solid curve: the \ion{Mg}{xii} spectroheliograph; blue dotted curve: EIT 171~\AA; green dashed curve: EIT 195~\AA; red dash dotted curve: EIT 284~\AA.}
\label{F:GOFT}
\end{figure}

The \ion{Mg}{xii} spectroheliograph \citep{zhi03a} onboard CORONAS-F/SPIRIT \citep{ora02, Zhitnik2002} obtained monochromatic images of the solar corona in the \ion{Mg}{xii} 8.42 \AA\ line. This line emits only at temperatures higher than 4 MK. The \ion{Mg}{xii} images differ from images of other ``hot imagers'' (like AIA or XRT); they do not contain a solar limb or any other low-temperature background. For a comparison of the \ion{Mg}{xii} images with other ``hot imagers'', see \citet{Reva2012, Reva2015}. Temperature-response function of the \ion{Mg}{xii} spectroheliograph is shown in Figure~\ref{F:GOFT}. 

We studied the period from 18\,--\,28 February 2002. At this time, the \ion{Mg}{xii} spectroheliograph worked with a 105-second cadence almost without data gaps. In these observations, the \ion{Mg}{xii} spectroheliograph registered binned images with a spatial resolution of 8$^{\prime\prime}$ and a 37~s exposure time.

The \ion{Mg}{xii} data were preprocessed: we subtracted the bias and dark-current frames. After preprocessing, the accuracy of the zero (average pixel count in the areas without hot objects) was $\approx$\,0.5\,DNs, and the value of the noise was $\approx$\,6\,DNs. The main source of the noise was an electronic interference. In binned \ion{Mg}{xii} images, a single photon causes $\approx$\,7\,DNs.

The EIT telescope onboard the {\it Solar and Heliospheric Observatory (SOHO)} spacecraft \citep{Domingo1995} took solar images at the wavelengths centered at 171, 195, 285, and 304~\AA. In a synoptic mode, EIT took images in all four channels every six hours; in the ``CME watch'' mode, the telescope took images in the 195~\AA\ channel every 12~minutes. The pixel size of the telescope is $2.6^{\prime\prime}$, and the spatial resolution is $5^{\prime\prime}$. The temperature response functions of the EIT 171, 195, and 284 channels are shown in Figure~\ref{F:GOFT}. EIT data were preprocessed with the standard {\sf eit\_prep.pro} procedure from the Solar Software package.

\section{Results}

\subsection{Hot Plasma on the Sun}

\begin{figure}[t]
\centering
\includegraphics[width = \textwidth]{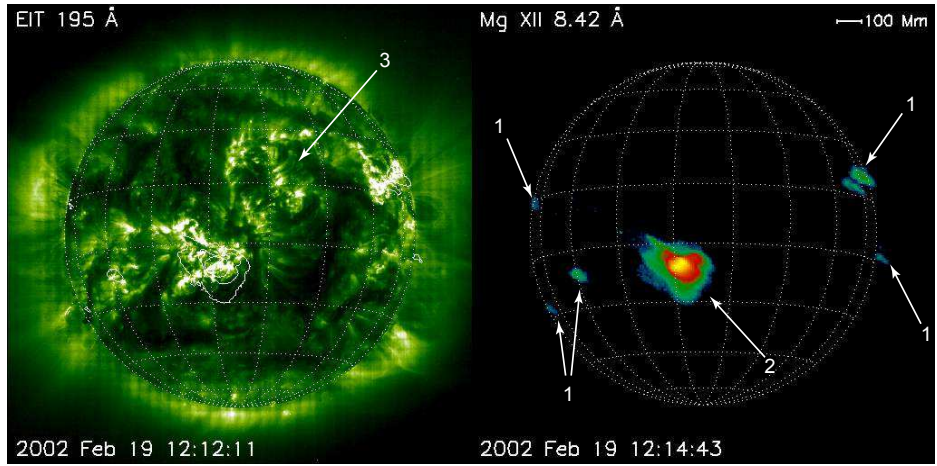}
\caption{Hot plasma observed by the \ion{Mg}{xii} spectroheliograph. Left: the EIT image; right: the \ion{Mg}{xii} spectroheliograph image (\textit{blue and green} correspond to low intensities, \textit{red and yellow} to high intensities). 1) small flare-like objects; 2) flaring active region with hot plasma; 3) non-flaring active region without hot plasma. (An Electronic Supplementary Material of this figure is available.)}
\label{F:Mg_EIT}
\end{figure}

\begin{figure}[t]
\centering
\includegraphics[width = \textwidth]{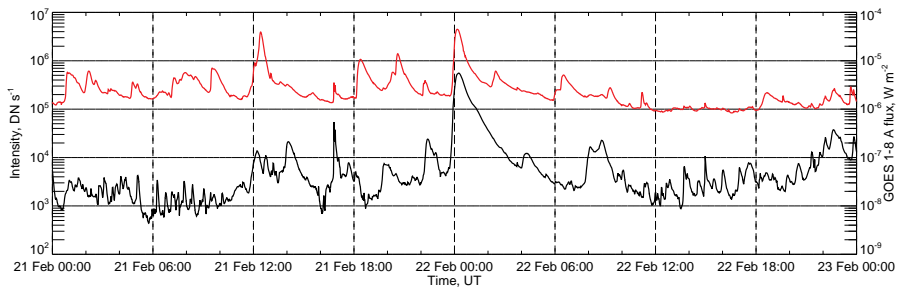}
\caption{Black: the \ion{Mg}{xii} spectroheliograph lightcurve of the flaring active region NOAA 09830; red: GOES 1\,--\,8\,\AA\ flux. This active region is marked with ``2'' in Figure~\ref{F:Mg_EIT}.}
\label{F:lightcurve}
\end{figure}

The Electronic Supplementary Material movie of Figure~\ref{F:Mg_EIT} shows the hot-plasma dynamics observed by the \ion{Mg}{xii} spectroheliograph. Only two types of hot objects were present on the Sun: the first is small isolated flare-like phenomena \citep{Reva2012}. During the period of observations, they occur at a rate of 20 per day. The second is large hot structures inside active regions. These are produced during flares or sequences of microflares. Their X-ray emission is highly variable (see Figure~\ref{F:lightcurve}). After the flare ends, these large structures fade away.

Except for rare microflares, there was no hot plasma in non-flaring active regions. Below we discuss this fact and try to estimate the upper limit of the hot-plasma emission measure and nanoflare frequency, which may be consistent with this lack of signal.

\subsection{Upper Limit on Hot Plasma in Non-Flaring Active Regions}

Although we do not observe hot plasma in non-flaring active regions, it is possible that its emission is so faint that the \ion{Mg}{xii} spectroheliograph cannot detect it. Nonetheless, we can estimate the upper limit on the amount of hot plasma using the sensitivity threshold of the instrument. 

For the estimate, we assume that the prediction of the nanoflare-heating model is correct: every pixel of the non-flaring active region has a small amount of hot plasma. We do not observe it, because its emission in the \ion{Mg}{xii} line [$I_\mathrm{hot}$] is lower than the minimal signal that the \ion{Mg}{xii} spectroheliograph can detect [$I_\mathrm{min}$].

$I_\mathrm{hot}$ is expressed as
\begin{equation}
I_\mathrm{hot} = EM_\mathrm{hot} G(T),
\end{equation}
where $EM_\mathrm{hot}$ is the column emission measure of the hot plasma, and $G(T)$ is the temperature response function of the \ion{Mg}{xii} spectroheliograph.

Hence, the upper limit on the column emission measure of hot plasma in non-flaring active regions is
\begin{equation}
EM_\mathrm{hot} = \frac{I_\mathrm{hot}}{G(T)} \le \frac{I_\mathrm{min}}{G(T)}.
\label{E:em_hot}
\end{equation}

\begin{figure}[t]
\centering
\includegraphics[width = 0.70\textwidth]{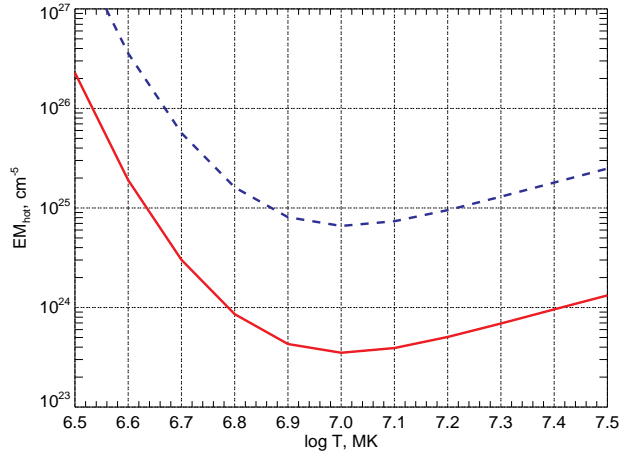}
\caption{Upper limit on the column emission measure of hot plasma in non-flaring active regions. Red solid curve: upper limit derived from the noise of the \ion{Mg}{xii} spectroheliograph; blue dashed curve: upper limit derived from the average pixel count.}
\label{F:EM}
\end{figure}

Two factors determine $I_\mathrm{min}$. The first one is the noise of the instrument: we cannot detect the signal that is lower than the noise. The second one is the photon statistics: we cannot detect less than one photon. $I_\mathrm{min}$ is the largest of the instrument noise and counts caused by a single photon.

\ion{Mg}{xii} images with a 37-second exposure had a noise of $N \approx$~6\,DNs. In binned \ion{Mg}{xii} images, a single photon caused $I_\mathrm{phot} \approx$~7\,DNs. Since $I_\mathrm{phot} > N$, then $I_\mathrm{min} = I_\mathrm{phot}$. We put the value of $I_\mathrm{phot}$ into Equation \ref{E:em_hot} and calculated the upper limit on the $EM_\mathrm{hot}$ (see Figure~\ref{F:EM}, blue). The column emission measure of hot plasma ($T \geq$~5\,MK) should not exceed $5 \times 10^{25}$\,cm$^{-5}$.

We made this estimate based on the absence of the signal in a single CCD pixel. However, each pixel of the non-flaring active region should contain hot plasma. If we integrate the signal in all pixels of the active region, the photon statistics will improve and the influence of the noise will decrease. This procedure could make our estimate better constrained. 

For the analysis, we chose the active region NOAA 09833 (see Figure~\ref{F:mg_eit_panel}). Except for a few microflares, there were no flares in this active region. The size of the area is $83 \times 90$ pixels. We averaged the flux of the chosen active region in the \ion{Mg}{xii} line. The average pixel count amounted to $-0.4$~DNs.

During averaging, the noise should decrease by a factor of $\sqrt{K}$ ($K = 83 \times 90$ is the number of pixels) and amount to $\approx$~0.05\,DNs. The minimal average detectable signal caused by photon statistics should be one photon divided by the number of pixel ($\approx$~0.001\,DNs). The absolute value of the average pixel count coincides with the error of the preprocessing.

As we see, after averaging, the noise and the photon statistics do not limit the sensitivity. The accuracy of the preprocessing determine the sensitivity threshold. We will use it as an upper limit on the signal from the hot plasma.

We put the  absolute value of the average pixel count into Equation \ref{E:em_hot} and calculated the upper limit on the column emission measure (see Figure~\ref{F:EM}, red). The column emission measure of hot plasma ($T \geq$~5~MK) should not exceed $3 \times 10^{24}$\,cm$^{-5}$. 

\subsection{DEM of the Active Region}

\begin{figure*}[t]
\centering
\includegraphics[width = \textwidth]{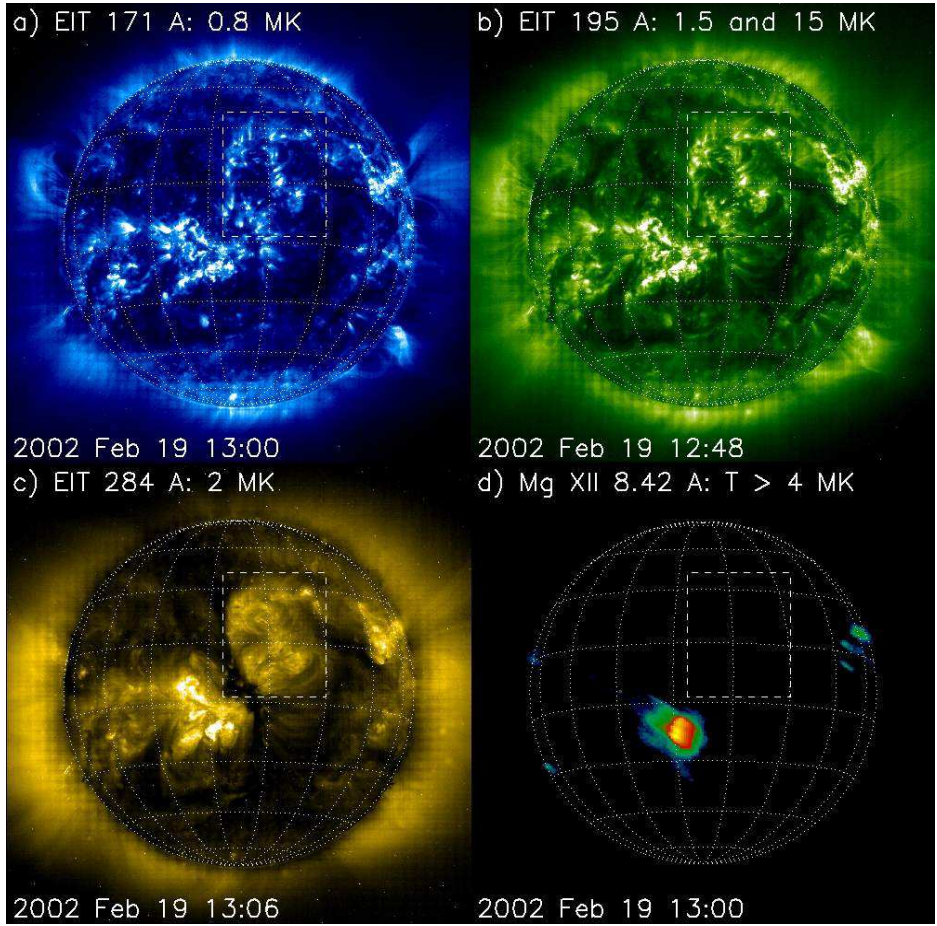}
\caption{Non-flaring active region NOAA 09833 that was used to determine DEM. a) EIT 171~\AA\ image; b) EIT 195~\AA\ image; c) EIT 284~\AA\ image; d) spectroheliograph \ion{Mg}{xii} image. Rectangle marks the active region that was used for DEM measurements.}
\label{F:mg_eit_panel}
\end{figure*}

To estimate the relative amount of the hot and warm (main temperature component) plasma, we reconstructed the DEM of the non-flaring active region NOAA 09833 (see Figure~\ref{F:mg_eit_panel}). For the selected active region, we extracted  its fluxes from the \ion{Mg}{xii}, EIT 171, 195, and 284 \AA\ channels. During extraction, we summed the signal inside the rectangle marked in the Figure~\ref{F:mg_eit_panel}. For the EIT channels, we subtracted from the fluxes the average flux of the quiet Sun multiplied by the area of the rectangle. The flux of the quiet Sun was calculated as the mean intensity on the boundary of the rectangle.

Using these four fluxes, we reconstructed the DEM with a genetic algorithm. To find a solution that fits the experimental fluxes, the method minimizes $\chi^2$ by mimicking the process of natural selection. The algorithm is described in Appendix~\ref{A:GA}.

Due to its random nature, the genetic algorithm returns different DEMs on different runs. Each of these solution satisfy the experimental fluxes equally well. If we run the algorithm multiple times, the spread of the solutions will be an estimate of the reconstruction accuracy. The accuracy is determined not by the method or measurement errors, but by the data set.

\begin{figure*}[t]
\centering
\includegraphics[width = \textwidth]{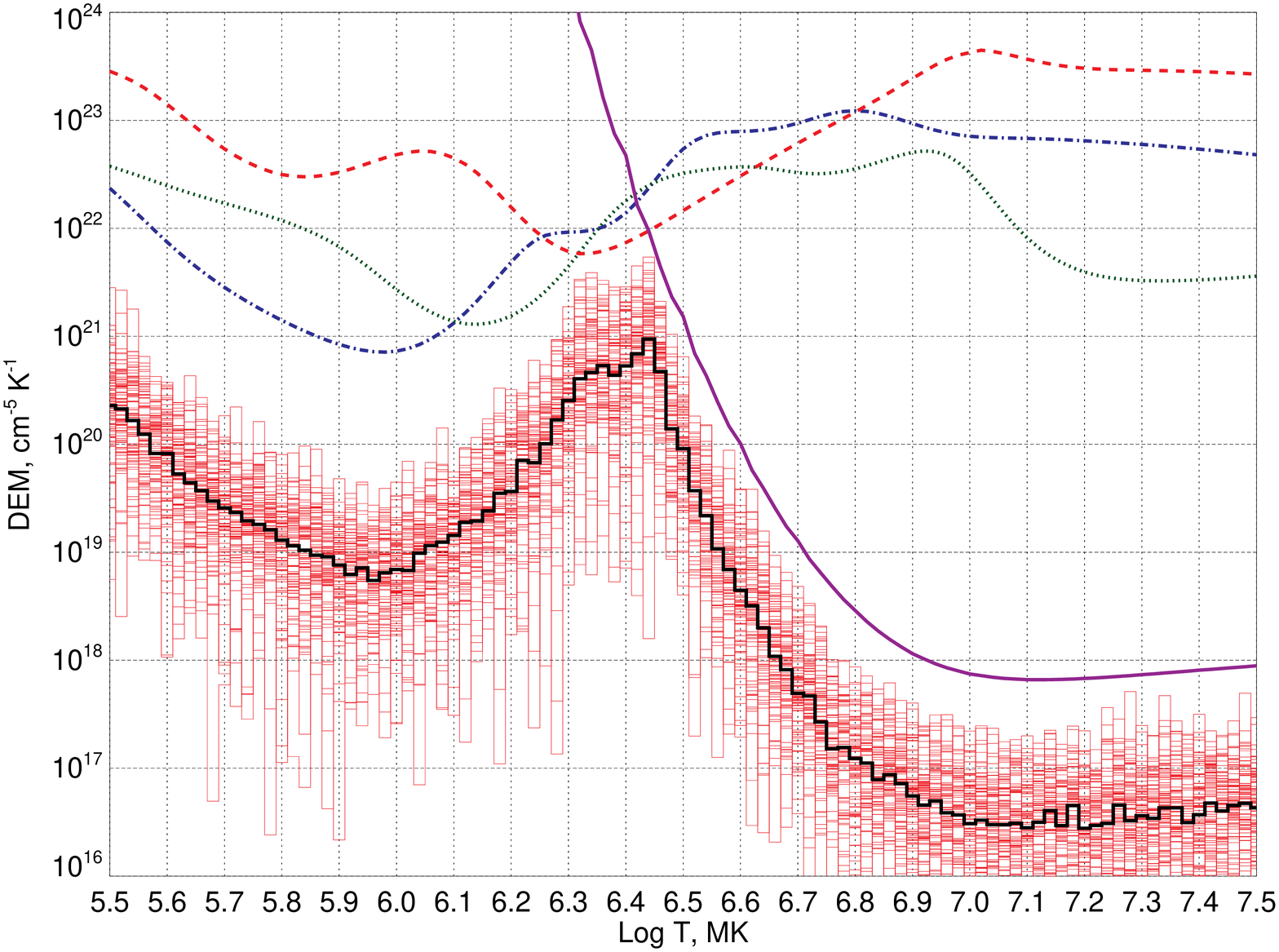}
\caption{DEM of the non-flaring active region marked in Figure~\ref{F:mg_eit_panel}. Red: DEMs obtained during different runs of genetic algorithm; black: their median. Blue dashed dotted DEM-loci: EIT 171~\AA; green dotted DEM-loci: EIT 195~\AA; red dashed DEM-loci: EIT 284~\AA; purple solid DEM-loci: \ion{Mg}{xii} spectroheliograph.}
\label{F:AR_DEM}
\end{figure*}

We ran the algorithm 100 times and plotted the result in the Figure~\ref{F:AR_DEM}. The amount of plasma with $T$~=~5\,MK ($\log T$~=~6.7) is four orders of magnitude lower than the main temperature component, and the amount of plasma with $T$~=~10\,MK is four to five orders lower.

At $\log T \ge 7.0$, only the \ion{Mg}{xii} flux constrains the DEM. The DEM in Figure~\ref{F:AR_DEM} at $\log T \ge 7.0$ shows the values that could be added to the DEM without increasing the \ion{Mg}{xii} flux to a level greater than the noise. These values are the DEM upper limit.

The \ion{Mg}{xii} spectroheliograph and EIT are not cross calibrated. The uncertainties in the cross calibration could affect the DEM reconstruction: the purple curve on Figure~\ref{F:AR_DEM} could move up or down. Below we will estimate the uncertainties in the cross calibration.

During large flares the \ion{Mg}{xii} flux is proportional to the GOES flux \citep{Urnov07}. Using this fact, the \ion{Mg}{xii} spectroheliograph was cross calibrated with GOES. The accuracy of the correlation between GOES and \ion{Mg}{xii} is $\approx$~10\,\% \citep{Urnov07}. The accuracy of the GOES calibration is $\approx$~30\,\% \citep{White2005, Viereck2017}. During cross calibration the CHIANTI atomic database \citep{Dere1997} was used, which has an accuracy of $\approx$~20\,\% \citep{DelZanna2011b}. This gives us a total accuracy of the \ion{Mg}{xii} spectroheliograph calibration of $\approx$~60\,\%.

Depending on the channel, the precision of the EIT calibration varies from 60 to 150\,\% \citep{Dere2000}. For the sake of the estimate, we will use the mean value of 100\,\%.

The total uncertainties in the cross calibration should be a factor of three. Although this is a high value, it is lower then the errors of the DEM reconstruction (factor of ten). Therefore, in this work, we neglect the uncertainties of the cross-calibration.

\section{Discussion}

\subsection{Nanoflare Frequency}

\begin{figure*}[t]
\centering
\includegraphics[width = 0.75\textwidth]{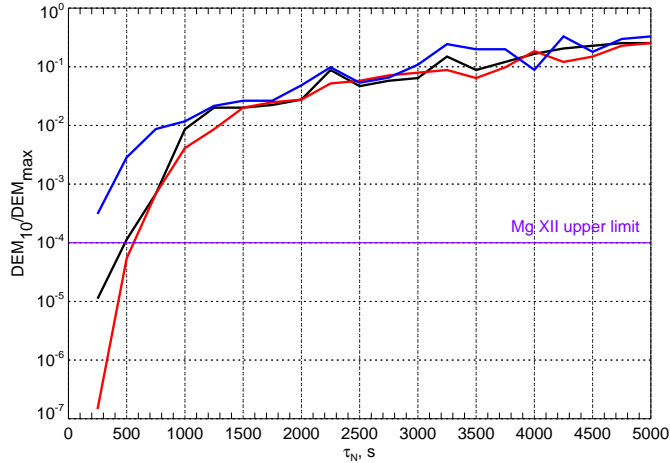}
\caption{Ratio of the $DEM_\mathrm{10}$ to the $DEM_\mathrm{max}$ as a function of the average time between nanoflares [$\tau_\mathrm{N}$]. The data were taken from the simulation performed by \citet{Cargill2014}. Black line denotes the simulations where nanoflares have power-law energy distribution with a slope $m = -2.5$ and the delay between nanoflares is proportional to their energy. Red line denotes the simulations where nanoflares have power-law energy distribution with a slope $m = -1.5$ and the delay between nanoflares is proportional to their energy. Blue line denotes the simulations where nanoflares have power-law energy distribution with a slope $m = -2.5$ and the delay between nanoflares is fixed. Purple line denotes the \ion{Mg}{xii} upper limit on the ratio of the DEMs of the hot and warm components.}
\label{F:ratio}
\end{figure*}

In this section, we will compare our results with the results obtained in the numerical simulations of nanoflare heating.

\citet{Cargill2014} performed numerical simulations of how a coronal loop should react to the sequence of nanoflares depending on the nanoflare frequency. The author found that for a low nanoflare frequency, the DEM has a hot component. The hot component vanishes for a high nanoflare frequency. 

\citet{Cargill2014} performed simulations for the following scenarios:

\begin{itemize}
\item nanoflares with power-law distribution of energies (slope $m = -2.5$) with a fixed delay between nanoflares;
\item nanoflares with power-law distribution of energies (slope $m = -2.5$) with a delay between nanoflares that is proportional to the nanoflare energy;
\item nanoflares with power-law distribution of energies (slope $m = -1.5$) with a delay between nanoflares that is proportional to the nanoflare energy.
\end{itemize}

For each of these regimes and different delays between nanoflares [$\tau_\mathrm{N}$], \citet{Cargill2014} provided plots of the active region DEMs. For each of these plots, we manually measured the ratio  of the DEM at 10~MK [$DEM_\mathrm{10}$] to the DEM of the main temperature component [$DEM_\mathrm{max}$]. Then we used the obtained values to build a plot of how this ratio depends on the delay between the nanoflares (see Figure~\ref{F:ratio}).

As we see from Figure~\ref{F:ratio}, the relative amount of hot plasma rapidly diminishes with the decrease of the delay between nanoflares. The \ion{Mg}{xii} data shows that this ratio should be less than 10$^{-4}$. Therefore, the delay between the nanoflares should be less than 500\,seconds. 

We emphasize that the parameters of the active region picked for the DEM calculation in our work and the one picked for simulations in \citet{Cargill2014} are different. Furthermore, the estimate of hot plasma based on the \ion{Mg}{xii} data is only an upper limit that is accurate within an order of magnitude.  Therefore, the value of the upper limit on the $\tau_\mathrm{N}$ should be considered as a rough estimate.

There is another way to explain the absence of emission in the \ion{Mg}{xii} images. If the heating impulse is very short, then the Mg ions could fail to reach the high-temperature ionization state before the electron temperature drops \citep{Reale2008, Bradshaw2009}. In this case, the hot emission will be absent not due to the low temperature, but due to the absence of the \ion{Mg}{xii} ions.

\subsection{Comparison with Other Observations}

\begin{table}[htb]
\centering
\caption{Observations of the hot plasma in non-flaring active regions. $DEM_\mathrm{X}$ is a differential emission measure at temperature $X$~MK.}
\begin{tabular}{
p{0.23\linewidth}    p{0.17\linewidth}      p{0.14\linewidth}         p{0.11\linewidth}                         p{0.17\linewidth}         }
\hline 
Work                 & Instrument           & Wavelength               & $\frac{DEM_\mathrm{5}}{DEM_\mathrm{3}}$ & $\frac{DEM_\mathrm{10}}{DEM_\mathrm{3}}$\\
\hline
\citet{Parenti2017}  & SUMMER               & \ion{Fe}{xix} 1118\,\AA  & $\leq$~0.1\,\%                          & $\leq$~0.1\,\%             \\
\citet{DelZanna2014} & FCS/SSM              & 13-20\,\AA               & $\leq$~1\,\%                            & N/A                        \\
\citet{Hannah2016}   & NuSTAR               & 2\,--\,78\,keV           & $\leq$~10\,\%                           & $\leq$~0.1\,\%             \\
\citet{Ishikawa2014} & FOXSI                & 6\,--\,8\,keV            & $\leq$~3\,\%                            & $\leq$~0.003\,\%           \\
\citet{Miceli2012}   & SphinX               & 1.34\,--\,7\,keV         & N/A                                     & N/A                        \\
\citet{Brosius2014}  & EUNIS-13             & \ion{Fe}{xix} 592.2\,\AA & N/A                                     & $\leq$~7.6\,\%             \\
\citet{Sylwester2010}& RESIK                & 3.4\,--\,6.1\,\AA        & $\leq$~0.1\,\%                          & N/A                        \\
\citet{Winebarger2012} & XRT                & 2\,--\,40\,\AA           & $\leq$~10\,\%                           & $\leq$~10\,\%              \\
\citet{Warren2012}   & AIA                  & 94\,\AA                  & $\leq$~1\,--\,10\,\%                    & $\leq$~1\,--\,10\,\%       \\
This work            & \ion{Mg}{xii}/SPIRIT & \ion{Mg}{xii} 8.42\,\AA  & $\leq$~0.01\,\%                         & $\leq$~0.001\,--\,0.01\,\% \\
\hline
\end{tabular}
\label{T:Observations}
\end{table}

In this section, we will compare the results of our observations with the observations mentioned in the Introduction. For comparison, we listed in Table~\ref{T:Observations} the values of the DEM ratio of the hot and warm components of the active region. If the work did not have these values, we tried to recalculate them from the numbers listed in the articles.

Our observations and the observations listed in Table~\ref{T:Observations} exhibit a similar picture. Active regions are heated up to average temperatures around 3\,MK. The steady hot plasma emission in non-flaring active regions is so faint that even the most sensitive instruments do not observe it. Heating above 3\,MK is observed only during flares or microflares \citep{Watanabe1992, Sterling1997}.

The \ion{Mg}{xii} limit on the relative amount of the hot plasma is at least one order of magnitude lower than the limit obtained in previous works. Only the data obtained during the FOXSI mission have a comparable limit \citep{Ishikawa2014}.

The advantages of the \ion{Mg}{xii} data presented in this work are temperature selectivity and the continuous observations. The \ion{Mg}{xii} data are monochromatic images of hot plasma that impose a strict constrains on the amount of hot plasma. Furthermore, the presented data are not just a few snapshots of the Sun; they represent thousands of images taken with a relatively high cadence. The observations shows that hot plasma is systematically not observed in the non-flaring active regions.

Although the \ion{Mg}{xii} data strictly constrained the amount of the hot plasma in non-flaring active regions, these observations are very old. Future instruments with better technology could improve the result.

\subsection{Hot Plasma in Flaring Active Regions}

Figure~\ref{F:lightcurve} shows that in flaring active regions the \ion{Mg}{xii} emission does not go to zero between spikes. This emission corresponds to 4~DNs per pixel, which is higher then the accuracy of the zero (0.5~DNs), but lower than the noise (6~DNs).

It is possible that this faint emission is caused by the nanoflares. However, it is also possible that this emission is the result of flares or microflares that previously occurred in the active region. Unfortunately, we do not have tools to distinguish these two cases.

\section{Conclusion}

In this work, we searched for hot plasma in the non-flaring active regions using the \ion{Mg}{xii} spectroheliograph onboard CORONAS-F/SPIRIT. This instrument built monochromatic images of the solar corona in the \ion{Mg}{xii} 8.42 \AA\ line, which emits only at temperatures higher than 4 MK. The \ion{Mg}{xii} images contain signal only from hot plasma without any low-temperature background.

Hot plasma was observed only in the flaring active regions or microflares. We did not observe any hot plasma in the non-flaring active regions. The hot plasma column emission measure in the non-flaring active region should not exceed $3 \times 10^{24}$\,cm$^{-5}$.

The hot DEM of the non-flaring active region is less than 0.01\,\% of the DEM of the main temperature component. Absence of the \ion{Mg}{xii} emission in the non-flaring active regions can be explained by weak and frequent nanoflares (delay less than 500\,seconds) or by very short and intense nanoflares that lead to non-equilibrium of ionization.

This work proves neither presence nor absence of hot plasma in non-flaring active regions. It neither confirms nor discards the nanoflare-heating model. This work estimates an upper limit on the relative amount of hot and warm plasma in non-flaring active regions and compares it with the predictions of the numerical simulations. This comparison limits the possible nanoflare frequency, which could help in further development of the nanoflare-heating model.

\begin{acks}
This research was supported by the Russian Science Foundation 
\newline
(project No. 17-12-01567).
\end{acks}

\section*{Disclosure of Potential Conflicts of Interest}
The authors declare that they have no conflicts of interest.

\appendix

\section{Genetic Algorithm}
\label{A:GA}

DEM reconstruction is a problem of calculating DEM based on the experimental fluxes obtained in different spectral channels. The fluxes and DEM are connected with the formula
\begin{equation}
I_i = \int G_i(T) DEM(T) \mathrm{d}T,
\label{E:GA_I}
\end{equation}
where $I_i$ is an experimental flux in the channel $i$, $G_i(T)$ is the temperature response function of the channel $i$, $DEM(T)$ is the differential emission measure, and $T$ is the temperature.

The genetic algorithm solves this problem by mimicking the process of the natural selection \citep{Siarkowski2008, Shestov2014}. It works in the following way:

\begin{enumerate}
\item The algorithm creates 1000 random DEMs.
\item For each of the DEM, the algorithm calculates the fluxes using Equation \ref{E:GA_I}.
\item Then the algorithm calculate a quantitative criteria of how well the calculated fluxes match with the experimental ones. As a quantitative criterion, we use $\chi^2$:
\begin{equation}
\chi^2 = \sum_i \frac{(C_i- I_i)^2}{\sigma_i^2},
\end{equation}
where $C_i$ is a calculated flux in the channel $i$, $I_i$ is an experimental flux in the channel $i$, and $\sigma_i$ is measurement error of the flux in the channel $i$.
\item The algorithm selects 100 DEMs with the best $\chi^2$. The rest of the DEMs are deleted.
\item The method creates 100 ``mutated'' DEMs: it copies the ``best'' DEMs and multiplies the value in each temperature bin by a random coefficient. This coefficient is uniformly distributed between 0.9 and 1.1.
\item The algorithm creates new 800 DEMs via ``breading''. This DEMs are calculated using the formula:
\begin{equation}
DEM = a DEM_1 + (1-a) DEM_2,
\end{equation}
where $DEM$ is the new DEM obtained via ``cross breading'', $DEM_1$ and $DEM_2$ are two DEMs randomly picked from the set of the ``best'' DEMs, and $a$ is a coefficient randomly picked from 0 to 1.
\item The algorithm repeats the procedure from the step ii) until the $\chi^2$ of the best DEM stops changing.
\item When the $\chi^2$ stops changing by more than 1\,\%, the algorithm returns the best DEM as a result.
\end{enumerate}

Equations \ref{E:GA_I} have an infinite number of solutions. A single run of the genetic algorithm randomly picks one solution. A second run will pick another solution, which will deviate from the first one. If we run the algorithm multiple times and plot the obtained solutions, we can estimate the uncertainties of the DEM reconstruction (see Figure~\ref{F:AR_DEM}).

The proportion of the ``best'', ``mutated'', and ``breading'' DEMs and the range of the coefficient that is used in the ``mutating'' procedure affect the speed of the calculation. We did not research which set of parameters is optimal. Most likely, there exists a better set of these parameters.

Sometimes the signal in some channels is below their sensitivity thresholds or a particular spectral line is blended with a stronger neighboring line. In this case, these channels provides only the upper limits on the fluxes. Equations \ref{E:GA_I} will change to:
\begin{equation}
I_i = \int G_i(T) DEM(T) dT
\end{equation}
\begin{equation}
U_j \geq \int G_j(T) DEM(T) dT,
\label{E:GA_U}
\end{equation}
where $I_i$ is the flux in the channel $i$, $G_i(T)$ is the temperature response function of the channel $i$, $DEM(T)$ is the differential emission measure, $T$ is the temperature, $U_j$ is the upper limit on the flux in the channel $j$, $G_j(T)$ is the temperature response function of the channel $j$.

Upper limits can enhance the DEM reconstruction. However, these limits cannot be treated like the usual fluxes. To use them, we need to modify the formula for calculating the $\chi^2$ to
\begin{equation}
\chi^2 = \sum_i \frac{(C_i- I_i)^2}{\sigma_i^2} + \sum_j F(C_j, U_j)
\end{equation}
where the function $F$ is defined as
\begin{equation}
F(C_j, U_j) = \left\{
                \begin{array}{lr}
                  \frac{(C_j- U_j)^2}{U_j^2}, & C_j > U_j\\
                  0,                          & C_j \leq U_j
                \end{array}
              \right.
\end{equation}
$F$ equals zero, when Equation \ref{E:GA_U} is satisfied, and $F$ approaches zero as $C_j$ approaches $U_j$.

\bibliographystyle{spr-mp-sola}
\bibliography{mybibl}

\end{article} 
\end{document}